\documentstyle[12pt]{article}
\date{}
\author{Valerii Dryuma,\\ Maxim Pavlov\\[5mm]
{\it Institute of Mathematics and Informatics, AS RM,}\\[3mm]
{\it 5 Academiei Street, 2028 Kishinev, Moldavia},
{\it MATI, Moskow, Russia} }
\title{ON  EQUATION FOR INITIAL VALUES IN THEORY OF THE
     SECOND ORDER ORDINARY DIFFERENTIAL EQUATIONS}

\newtheorem{pr}{Proposition}
\newtheorem{rem}{Remark}
\begin{document}
\maketitle

\begin{abstract}
\ \ \ \ We consider the properties of the
        second order nonlinear differential equations $b''= g(a,b,b')$
        with the function $g(a,b,b'=c)$ satisfying the following nonlinear partial
        differential equation
        $$
        g_{aacc}+2cg_{abcc}+2gg_{accc}+c^2g_{bbcc}+2cgg_{bccc}
        $$
        $$
        +g^2g_{cccc}+(g_a+cg_b)g_{ccc}-4g_{abc}-4cg_{bbc} -cg_{c}g_{bcc}
        $$
        $$
        -3gg_{bcc}-g_cg_{acc}+ 4g_cg_{bc}-3g_bg_{cc}+6g_{bb} =0\>.
        $$

        Any equation $b''=g(a,b,b')$ with this condition on function
        $g(a,b,b')$ has the General Integral $F(a,b,x,y)=0$
        shared with General Integral of the second order ODE's $y''=f(x,y,y')$
        with condition $\frac{\partial^4 f}{\partial y'^4}=0$
        on function $f(x,y,y')$ or
        $$
        y''+a_{1}(x,y)y'^3+3a_{2}(x,y)y'^2+3a_{3}(x,y)y'+a_{4}(x,y)=0
        $$
        with some coefficients $a_{i}(x,y)$.

\end{abstract}

\section{Introduction}

     The relation between the equations in form
\begin{equation}
  y''+a_{1}(x,y)y'^3+3a_{2}(x,y)y'^2+3a_{3}(x,y)y'+a_{4}(x,y)=0  \label{Cartan1}
\end{equation}
and
\begin{equation}
b''=g(a,b,b') \label{Dua1}
\end{equation}
 with function $g(a,b,b')$ satisfying  the p.d.e
$$
     g_{aacc}+2cg_{abcc}+2gg_{accc}+c^2g_{bbcc}+2cgg_{bccc}+
$$
\begin{equation}
    g^2g_{cccc}+(g_a+cg_b)g_{ccc}-4g_{abc}-4cg_{bbc} -cg_{c}g_{bcc}-
\end{equation}
$$
  3gg_{bcc}-g_cg_{acc}+ 4g_cg_{bc}-3g_bg_{cc}+6g_{bb} =0 \label{Dua2}.
$$
from geometrical point of view was studied by E.Cartan \cite{Cartan1}.

  In fact, according to the exp\-res\-sions on curvature of the
space of linear elements (x,y,y') connected with equation (1)
$$
\Omega^1_2=a[\omega^2 \wedge \omega^2_1]\,,\quad
\Omega^0_1=b[\omega^1 \wedge \omega^2]\,,\quad
\Omega^0_2=h[\omega^1 \wedge \omega^2]+k[\omega^2 \wedge \omega^2_1]\>.
$$
where:
$$
a=-\frac{1}{6}\frac{\partial^4 f}{\partial y'^4}\,,\quad h=\frac{\partial b}{\partial y'}\,, \quad
k=-\frac{\partial \mu}{\partial y'}-\frac{1}{6}\frac{\partial^2 f}{\partial^{2} y'}\frac{\partial^3 f}{\partial^{3} y'}\,,
$$
and
\begin{eqnarray*}
6b & = & f_{xxy'y'}+2y'f_{xyy'y'}+2ff_{xy'y'y'}+y'^2f_{yyy'y'}+2y'ff_{yy'y'y'} \\
   & + & f^2f_{y'y'y'y'}+(f_x+y'f_y)f_{y'y'y'}-4f_{xyy'}-4y'f_{yyy'} - y'f_{y'}f_{yy'y'}\\
   & - &  3ff_{yy'y'}-f_{y'}f_{xy'y'}+ 4f_{y'}f_{yy'}-3f_{y}f_{y'y'}+6f_{yy} \>.
\end{eqnarray*}
two types of equations by a natural
way are evolved: the first type from the con\-di\-tion $a =0$ and second type
from the condition $b =0$.

    The first condition $a=0$ the equation in form (1)
is detemined and the second condition lead to the equations (2)
where the function $g(a,b,b')$ satisfies the above p.d.e.~(3).

    From the elementary point of view the relation between both equations
(1) and (2) is a result of the special properties of their
General Integral
$$
F(x,y,a,b)=0.
$$
     So we have the
following fundamental diagramm:
$$
\begin{array}{ccccc}
 &  & F(x,y,a,b)=0 &  & \\
 & \swarrow \nearrow &  & \searrow \nwarrow & \\
y''=f(x,y,y') & & & & b''=g(a,b,b') \\
 & & & & \\
 \Updownarrow & & & & \Updownarrow \\
 & & & & \\
  M^3(x,y,y') & & \Longleftrightarrow & & N^3(a,b,b')  
\end{array}
$$
which  is  presented  the  General  Integral $F(x,y,a,b)=0$ 
 (as  some  3-dim  orbifold) in  form  of  the  twice
nontrivial  fibre bundles on circles over corresponding surfaces:
$$
M^{3}(x,y,y')= U^{2}(x,y) \times S^1 \quad
\mbox{\rm and}  \quad
N^{3}(a,b,b')= V^{2}(a,b) \times S^1\>.
$$

\section{An examples of solutions of dual equation}

    Let us consider  the solutions of equation ~(3).

   It has many types of
reductions and the symplest of them are
$$
 g=c^{\alpha}\omega[ac^{\alpha-1}],\quad g=c^{\alpha}\omega[bc^{\alpha-2}],
\quad g=c^{\alpha}\omega[ac^{\alpha-1},bc^{\alpha-2}],
\quad g=a^{-\alpha}\omega[ca^{\alpha-1}],
$$
$$
\quad g=b^{1-2\alpha}\omega[cb^{\alpha-1}],
\quad g=a^{-1}\omega(c-b/a),
\quad g=a^{-3}\omega[b/a,b-ac],\quad
 g=a^{\beta/\alpha-2}\omega[b^{\alpha}/a^{\beta},
c^{\alpha}/a^{\beta-\alpha}].
$$

    For any type of reduction we can write corresponding equation
~(2) and
then integrate it.

    As example for the function
$$
g=a^{-\gamma}A(ca^{\gamma-1})
$$
we get the equation
$$
[A+(\gamma-1)\xi]^2 A^{1V}+3(\gamma-2)[A+(\gamma-1)\xi]A^{111}+(2-\gamma)A^1A^{11}+
(\gamma^2-5\gamma+6)A^{11}=0.
$$

     One solution of this equation is
$$
A-(2-\gamma)[\xi(1+\xi^2)+(1+\xi^2)^{3/2}]+(1-\gamma)\xi
$$
  
    This solution is coresponded to the equation
$$
b''=\frac{1}{a}[b'(1+b'^2)+(1+b'^2)^{3/2}]
$$
with Gneral Integral
$$
F(x,y,a,b)=(y+b)^2+a^2-2ax=0
$$
The dual equation has the form
$$
y''=-\frac{1}{2x}(y'^3+y')
$$
 
   Remark that the first examples of
solutions of equation (3) was obtained in [6-9].

    The
\begin{pr}
 Equation ~(3) can be represent in form
\begin{eqnarray}
   g_{ac} +  gg_{cc} - g_{c}^{2}/2 + cg_{bc} -2 g_{b} = h(a,b,c),\\\nonumber
h_{ac}  +   gh_{cc} - g_{c}h_{c} + ch_{bc} -3h_{b} = 0.
\end{eqnarray}
\end{pr}

    From this is followed that exists the class of equations ~(\ref{Dua1})
 with  function $g(a,b,c)$  satisfying the condition
\begin{equation}
   g_{ac}  +  gg_{cc} - g_{c}^{2}/2 + cg_{bc} -2 g_{b} = 0. \label{H}
\end{equation}
which is more readily solved then equation ~(3).

    Here we present some solutions of the equation (8) as function
depending on two variables $g=g(a,c)$

   In case when $g=g(a,c)$ and $h=0$ we have the equation
$$
g_{ac}+gg_{cc}-\frac{1}{2}{g_c}^2=0\>.
$$

   To integrate this equation we can transform its in more convenient form using variable
$g_c=f(a,c)$. Then one obtains:
$$
2f_cf_{ac}+(f^2-2f_a)f_{cc}=0\>.
$$
After the Legendre-transformation we obtain the equation:
$$
[(\xi\omega_{\xi}+\eta\omega_{\eta}-\omega)^2-
2\xi]{\omega}_{\xi\xi}-2\eta\omega_{\xi\eta}=0\>.
$$
Using the new variable $\xi\omega_{\xi}+\eta\omega_{\eta}-\omega=R$ we have
the new equation for $R$:
$$
R_{\xi}-\frac{1}{2}R^2\omega_{{\xi}{\xi}}=0\>
$$
and the following relations:
$$
\omega_{\eta}=\frac{\omega}{\eta}+\frac{R}{\eta}+\frac{2{\xi}}{{\eta}R}-\frac{{\xi}A(\eta)}{\eta}\>,
$$
$$
\omega_{\xi}=-\frac{2}{R}+A(\eta)\>
$$
with arbitrary function $A(\eta)$.
    From the conditions of compatibility is followed:
$$
2{\eta}R_{\eta}+R_{\xi}(2\xi-R^2)+\eta A_{\eta}R^2=0\>.
$$
Integrating this equation we can obtain general integral.

 In the particular case:
$A=\frac{1}{\eta}$ we have:
$$
\frac{R^2}{R-2\eta}=-\frac{\xi}{\eta}+\Phi(\frac{1}{\eta}-\frac{2}{R})\>.
$$
At the condition $A=0$ we obtain the equation:
$$
2{\eta}R_{\eta}+(2\xi-R^2)R_{\xi}=0\>,
$$
which has the solution:
$$
R^2=2{\xi}+2{\eta}{\Phi(R)}\>,
$$
were $\Phi(R)$ is arbitrary function.

     After  choosing the function $\Phi(R)$ we can find the function
$\omega$ and then using the inverse Legendre transformation  the function
$g$ which is determined dual equation $b''=g(a,c)$.

\begin{rem}
   The solutions of the equations of type 
$$
u_{xy}=uu_{xx}+\epsilon u_{x}^2
$$
was constructed in [19]. In  work of [20] was showed that they 
can be present in form
$$
u=B'(y)+\int[A(z)-\epsilon y]^{(1-\epsilon)/ \epsilon} dz,
$$
$$
x=-B(y)+\int[A(z)-\epsilon y]^{1/ \epsilon} dz.
$$

    To integrate above equations we apply the parametric representation
$$
g=A(a)+U(a,\tau), \quad c=B(a)+V(a,\tau).\eqno(11)
$$
Using the formulas
$$
g_c=\frac{g_{\tau}}{c_{\tau}}, \quad g_{a}=g_{a}+g_{\tau}\tau_{a}
$$
we get after the substitution in (10) the conditions
$$
A(a)=\frac{d B}{d a}
$$
and 
$$
U_{a \tau}-\left(\frac{V_{a} U_{\tau}}{V_{\tau}}\right)_{\tau}+
U \left(\frac{ U_{\tau}}{V_{\tau}} \right)_{\tau} -
\frac{1}{2} \frac{U_{\tau}^2}{V_{\tau}}=0.
$$

     So we get one equation for two functions $U(a,\tau)$ and $V(a,\tau)$.
Any solution of this equation are determined  the solution  of 
equation (10) in form (11).
    
     Let us consider the examples.
$$
A=B=0, \quad U=2\tau-\frac{a\tau^2}{2}, \quad V=a\tau-2\ln(\tau)
$$

   Using the representation
$$
U=\tau \omega_{\tau}-\omega,\quad V=\omega_{\tau}
$$
it is possible to obtain others solutions of this equation.

Equation
$$
g_{ac}=gg_{cc}-\frac{1}{2}{g_c}^2\>.
$$
can be integrate in explicite form and solutions are
$$
g=-H'(a)+\int\frac{dz}{[A(z)+\frac{1}{2}a]^3}\>,
$$
$$
c=H(a)+\int\frac{dz}{[A(z)+\frac{1}{2}a]^2}\>,
$$
with arbitrary functions $H(a)$ and $A(z)$.

   In fact, for $A(z)=z$
we have
$$
g=-H'(a)+\int\frac{dz}{[z)+\frac{1}{2}a]^3}=-H'(a)-\frac{1}{2}\frac{1}
{[z)+\frac{1}{2}a]^2}\>,
$$
and
$$
c=H(a)+\int\frac{dz}{[z+\frac{1}{2}a]^2}=H(a)-
\frac{1}{[z)+\frac{1}{2}a]^3}\>,
$$

As result we get the solution
\end{rem}

\begin{rem}
In general case the equation
$$
g_{acc}+gg_{ccc}=0,
$$
is equivalent the equation
$$
g_{ac}+gg_{cc}-\frac{1}{2}{g_c}^2=B(a)\>.
$$

    It can be intgrate with help of Legender- transformation
as in previous case.

Realy, we get
$$
[(\xi\omega_{\xi}+\eta\omega_{\eta}-\omega)^2-
2\xi+2B(\omega_{\xi})]{\omega}_{\xi\xi}-2\eta\omega_{\xi\eta}=0\>
$$
and the relation
$$
2R_{\xi}=[R^2+2B(\omega_{\xi})\omega_{\xi \xi}.
$$
It can be written in form
$$
2\frac{dR}{d\Omega}=R^2+2B(\Omega)
$$
using the notation
$$
\omega_{\xi}=\Omega
$$

\end{rem}

\begin{pr}

  In case $h\neq 0$ and $g=g(a,c)$ the system ~(3) is equivalent the equation
\begin{equation}
\Theta_a(\frac{\Theta_a}{\Theta_c})_{ccc}-
\Theta_c(\frac{\Theta_a}{\Theta_c})_{acc}=1
\end{equation}
where
$$
g=-\frac{\Theta_a}{\Theta_c}\quad h_c=\frac{1}{\Theta_c}
$$
\end{pr}

    To integrate this equation we use the presentation
$$
c=\Omega(\Theta,a)
$$

    From the relations
$$
1=\Omega_{\Theta}\Theta_c, \quad 0=\Omega_{\Theta}\Theta_a+\Omega_c
$$
we get
$$
\Theta_c=\frac{1}{\Omega_{\Theta}},\quad 
\Theta_a=-\frac{\Omega_a}{\Omega_{\Theta}}
$$
and
$$
\frac{\Omega_a}{\Omega_{\Theta}}(\Omega_a)_{ccc}+
\frac{1}{\Omega_{\Theta}}(\Omega_a)_{cca}=1
$$

    Now we get
$$
\Omega_{ac}=\frac{\Omega_{a \Theta}}{\Omega_{\Theta}}=
(\ln \Omega_{\Theta})_a=K,\quad
\Omega_{acc}=\frac{K_{\Theta}}{\Omega_{\Theta}},\quad
$$
$$
\Omega_{accc}=(\frac{K_{\Theta}}{\Omega_{\Theta}})_{\Theta}\frac{1}{\Omega_{\Theta}},\quad
(\Omega_{acc})_a=
(\frac{K_{\Theta}}{\Omega_{\Theta}})_a-\frac{\Omega_a}{\Omega_{\Theta}}
(\frac{K_{\Theta}}{\Omega_{\Theta}})_{\Theta}
$$

     As result the equation (6) take the form
\begin{equation}
\left[\frac{(\ln\Omega_{\Theta})_{a\Theta}}{\Omega_{\Theta}}\right]_a=\Omega_{\Theta}
\end{equation}
and can be integrate under the substitution
$$
\Omega(\Theta,a)=\Lambda_a
$$

   So we get the equation
\begin{equation}
\Lambda_{\Theta\Theta}=\frac{1}{6}\Lambda_{\Theta}^3+
\alpha(\Theta)\Lambda_{\Theta}^2+
\beta(\Theta)\Lambda(\Theta)+\gamma(\Theta)
\end{equation}
with arbitrary coefficients $\alpha,\beta,\gamma$.
     
    This is Abel's type of equation
$$
y'=A(x)y^3+B(x)y^2+C(x)y+D(x) 
$$
  
 It can be rewriten in form
$$
y'=A(y-\phi)^3+\theta(y-\phi)^2+\lambda(y-\phi)+\phi'
$$
or
$$   
z'=A z^3+\theta z^2+\lambda z
$$  
Let us considere the examples.

     1. $\alpha=\beta=\gamma=0$

    The solution of equation (8) is
$$
\Lambda=A(a)-6\sqrt{B(a)-\frac{1}{3}\Theta}
$$
and we get
$$
c=A'-\frac{3B'}{\sqrt{B-\frac{1}{3}\Theta}}
$$
or
$$
\Theta=3B-27\frac{B'^2}{(c-A')^2}
$$
This solution is corresponded to the equation
$$
b''=-\frac{\Theta_a}{\Theta_c}=-\frac{1}{18B'}b'^3+\frac{A'}{6B'}b'^2+
(\frac{B''}{B'}-\frac{A'^2}{6B'})b'+A''+\frac{A'^3}{18B'}-\frac{A'B''}{B'}
$$
cubical on the first derivatives $b'$ with arbitrary coefficients $A(a),B(a)$.
This equation is equivalent to the equation 
$$
b''=0
$$
under the point transformation.

   The following example is the solution of equation (8) in form
$$
   g=b^{1-2\alpha}\omega[cb^{\alpha-1}]
$$
Under this reduction one obtains the equation on the function $\omega(\xi=
cb^{\alpha-1})$
$$
\omega\omega''-\frac{\omega'^2}{2}+(\alpha-1)\xi^2\omega''+
(2-3\alpha)\xi\omega'+2(2\alpha-1)\omega=0.
$$
        To make the new variable $\theta=\omega+(\alpha-1)\xi^2$ we obtain
$$
\theta\theta''-\frac{\theta'^2}{2}-\alpha\xi\theta'+2\alpha\theta=0.
$$
  This equation has solution in parametrical form
$$
\theta=\gamma\tau E(\tau), \quad \xi=\frac{\gamma E(\tau)}{\beta}
$$
where
\begin{equation}
E(\tau)=\exp[-\int \frac{\tau d\tau}{(\tau -1/2)^2 + \alpha/\beta^2-1/4}]
\end{equation}
where $\beta, \gamma$ are parametrs
and the explicit form of this integral depends on the value
$$
\epsilon=\alpha/\beta^2-1/4.
$$

   This solution is corresponed to the family of equations
$$
b''=b^{1-2\alpha}[\theta+(1-\alpha)\xi^2]
$$
and (1) forming dual paar.

  The values of coefficients $a_i(x,y,\alpha,\beta,\gamma)$ in corresponding
equation (1) can be change radicaly at the
variation  of  parameters as it is showed the calculation of integral (10).

\section{Acknowledgement}

     This work was supported by INTAS-93-0166.
The V.D is grateful to
the Physical Departement of Lecce University
for support and kind hospitality.

\end{document}